\documentclass[10pt,twocolumn,english,final,5p,times]{revtex4}
\usepackage[T1]{fontenc}
\usepackage[latin9]{inputenc}
\usepackage{amsmath}
\usepackage{amssymb}
\usepackage{graphicx}

\makeatletter
\@ifundefined{textcolor}{}
{%
 \definecolor{BLACK}{gray}{0}
 \definecolor{WHITE}{gray}{1}
 \definecolor{RED}{rgb}{1,0,0}
 \definecolor{GREEN}{rgb}{0,1,0}
 \definecolor{BLUE}{rgb}{0,0,1}
 \definecolor{CYAN}{cmyk}{1,0,0,0}
 \definecolor{MAGENTA}{cmyk}{0,1,0,0}
 \definecolor{YELLOW}{cmyk}{0,0,1,0}
}

\makeatother

\usepackage{babel}
\begin{document}

\title{Thermal entanglement in an orthogonal dimer-plaquette chain with alternating Ising-Heisenberg
coupling}

\author{H. G. Paulinelli, S. de Souza and Onofre Rojas}

\affiliation{Departamento de Ciencias Exatas, Universidade Federal de Lavras,
37200-000, Lavras - MG, Brazil }
\begin{abstract}

In this paper we explore the entanglement in orthogonal dimer-plaquette Ising-Heisenberg chain, assembled between
plaquette edges, also known as orthogonal dimer plaquettes. The quantum entanglement properties involving an infinite chain structure are quite important, not only because the mathematical calculation is cumbersome but also because real materials are well represented by infinite chain.
Using the local gauge symmetry of this model, we are able to map onto a simple
spin-1 like Ising and spin-1/2 Heisenberg dimer model with single
effective ion anisotropy.  Thereafter this model can be solved using
the decoration transformation and transfer matrix approach. First,
we discuss the phase diagram at zero temperature of this model, where
we find five ground states, one ferromagnetic, one antiferromagnetic,
one triplet-triplet disordered and one triplet-singlet disordered
phase, beside a dimer ferromagnetic-antiferromagnetic phase. In addition, we discuss the thermodynamic properties
such as entropy, where we display the residual entropy. Furthermore,
using the nearest site correlation function it is possible also to
analyze the pairwise thermal entanglement for both orthogonal dimers,
additionally we discuss the threshold temperature of the entangled
region as a function of Hamiltonian parameters. We find quite interesting thin reentrance threshold temperature for one of the dimers, and we also discuss the differences and similarities for both dimers.
\end{abstract}


\maketitle

\section{Introduction}
Recently, several theoretical investigation are dedicated   to quantum entanglement, which is one of the most fascinating
types of correlations that can be shared only among quantum systems\cite{AmicoHorod}.
In recent years, many efforts have been devoted to characterizing
qualitatively and quantitatively the entanglement properties of condensed
matter systems, which are the natural candidate for application in
quantum communication and quantum information. In this sense, it is
quite relevant to study the entanglement of solid state systems such
as spin chains\cite{qubit-Heisnb}. The Heisenberg chain is one of
the simplest quantum systems, which exhibits the entanglement, due
to the Heisenberg interaction is non localized in the spin system.
Several studies have been done on the threshold temperature for the
pairwise thermal entanglement in the Heisenberg model with a finite
number of qubits. Thermal entanglement of the isotropic Heisenberg
chain has been studied in the absence \cite{wang} and in the presence
of external magnetic field \cite{arnesen}. 

On the other hand, quasi two-dimensional magnets have been attracted since 90 decades,
such as the quasi-two-dimensional magnet $\mathrm{CaV}_{4}\mathrm{O}_{9}$\cite{tanigushi},
that has a layered structure where the magnetic $\mathrm{V}^{4+}$
ions have spin 1/2 and form a 1/5-depleted square lattice. As well
as the polycrystalline $\text{SrCu}_{2}\text{(BO}_{3})_{2}$ having
a two dimensional (2D) orthogonal network of Cu dimers, this cuprate,
provides a 2D spin-gap system in which the ground state can be solved
exactly\cite{miyahara,kageyama}. These quasi-two-dimensional systems
are topologically equivalent to the theoretical model proposed by
Shastry and Sutherland\cite{shastry}.

Motivated by the above real materials Ivanov and Richter\cite{ivanov}
proposed the class of one-dimensional Heisenberg spin models (plaquette
chains) related to the real materials\cite{tanigushi,miyahara,kageyama}, were analyzed the zero temperature
magnetic properties through numerical and analytical results  \cite{ivanov,richter}. While in reference \cite{schu}
was discussed the sequence of first-order quantum phase transitions
in a frustrated spin half dimer-plaquette chain. A detailed investigation
about the first-order quantum phase transition of the orthogonal-dimer
spin chain also was considered by Koga et al.\cite{koga}, as well
as the frustration-induced phase transitions in the spin-$S$ orthogonal-dimer
chain\cite{koga02}. 

A more recent investigation was developed by Ohanyan and Honecker\cite{vadim12},
where they have been discussed the magnetothermal properties of the
Ising-Heisenberg orthogonal-dimer chain with triangular XXZ clusters.
Furthermore, in the last decade several quasi-one-dimensional Ising-Heisenberg
model such as diamond chain were intensively investigated, mainly
the thermodynamic properties and geometric frustration\cite{strecka06,strecka09,valverde,vadim},
magneto-caloric effect\cite{pereira2}, as well as thermal entanglement\cite{spra,entg-ananik},
among other physical quantities. Some other variant of the Ising-Heisenberg
model also were considered, such as Ising-Hubbard model\cite{lisnii}
and Hubbard model in the quasi-atomic limit\cite{hubbard} besides
spinless electrons\cite{spinless} in diamond chain.

The outline of this work is as follows. In sec. 2 we present the dimer-plaquette
Ising-Heisenberg chain; it is also discussed the zero temperature
phase diagram. In sec. 3 we present the exact thermodynamic solution
of the model, thus we discuss the entropy, specific heat and correlation function. In sec.
4 we discuss the thermal entanglement and its threshold temperature.
Finally in sec. 5, we summarize our results and draw our conclusions.

\section{Orthogonal dimer-plaquette Ising-Heisenberg chain}

\begin{figure}
\centering{}\includegraphics[scale=0.4]{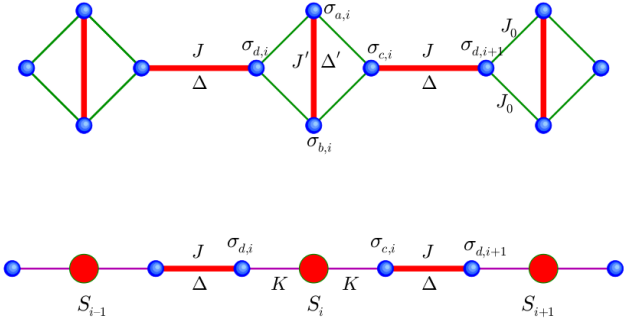}\caption{\label{fig:Schm-chain}(Color online) Schematic representation of
dimer-plaquette Ising-Heisenberg chain. (Above) Thick line correspond
to Heisenberg coupling, while the thin line corresponds to Ising coupling.
(Below) Dimer plaquette mapping through local gauge symmetry.}
\end{figure}

The theoretical investigation of the orthogonal dimer-plaquette models
are motivated not only from the theoretical point of view, but also
from the experimental viewpoint. It is worth to remark that the Heisenberg
orthogonal dimer-plaquette model cannot be solved exactly at finite
temperature. Driven by the comments given in the introduction,
we consider the orthogonal-dimer plaquette chain with Ising-Heisenberg
coupling as schematically described in figure \ref{fig:Schm-chain}.
Therefore, the Hamiltonian for an orthogonal dimer-plaquette chain
could be expressed by
\begin{alignat}{1}
H= & -\sum_{i=1}^{N}\left\{ J^{\prime}(\boldsymbol{\sigma}_{a,i},\boldsymbol{\sigma}_{b,i})_{\Delta^{\prime}}+J(\boldsymbol{\sigma}_{c,i},\boldsymbol{\sigma}_{d,i})_{\Delta}\right.\nonumber \\
 & \left.+J_{0}[(\sigma_{a,i}^{z}+\sigma_{b,i}^{z})\sigma_{c,i}^{z}+(\sigma_{a,i+1}^{z}+\sigma_{b,i+1}^{z})\sigma_{d,i}^{z}]\right\} ,\label{eq:Ham-orig}
\end{alignat}
with
\begin{equation}
J^{\prime}(\boldsymbol{\sigma}_{a,i},\boldsymbol{\sigma}_{b,i})_{\Delta^{\prime}}=J^{\prime}(\sigma_{a,i}^{x}\sigma_{b,i}^{x}+\sigma_{a,i}^{y}\sigma_{b,i}^{y})+\Delta^{\prime}\sigma_{a,i}^{z}\sigma_{b,i}^{z}
\end{equation}
where $\sigma_{\gamma,i}^{\alpha}$ are the spin operators also known
as Pauli matrices (with $\alpha=\{x,y,z\}$) at plaquette $i$ for
particles $\gamma=\{a,b,c,d\}$, for detail see figure \ref{fig:Schm-chain}.
The thick line correspond to Heisenberg coupling, while the thin line
corresponds to Ising coupling. The Ising coupling parameter is denoted
by $J_{0}$, whereas $J$ ($J^{\prime}$) represents the $x$ and
$y$ components of Heisenberg coupling and with $\Delta$ ($\Delta^{\prime}$)
we mean the anisotropic ($z$-component) coupling in the Heisenberg
term for $ab$-dimer and $cd$-dimer respectively.

To transform this model onto the well known mixed spin \textquotedbl{}Ising\textquotedbl{}--Heisenberg
model, we use the following definition $S_{i}^{\alpha}=\sigma_{a,i}^{\alpha}+\sigma_{b,i}^{\alpha}$.
By the use of $S_{i}^{\alpha}$ definition we obtain the identity:
$(S_{i}^{\alpha})^{2}=2+2\sigma_{a,i}^{\alpha}\sigma_{b,i}^{\alpha}$.
Thereafter, we can easily establish the following transformation
\begin{equation}
J^{\prime}(\boldsymbol{\sigma}_{a,i},\boldsymbol{\sigma}_{b,i})_{\Delta^{\prime}}=\frac{J^{\prime}}{2}\boldsymbol{S}_{i}^{2}+\frac{\Delta^{\prime}-J^{\prime}}{2}\left(S_{i}^{z}\right)^{2}-(2J^{\prime}+\Delta^{\prime}),\label{eq:spin-transf}
\end{equation}
and by $\boldsymbol{S}_{i}^{2}$ we just denote $\boldsymbol{S}_{i}^{2}=\boldsymbol{S}_{i}.\boldsymbol{S}_{i}$,
so, this matrix is given by
\begin{equation}
\boldsymbol{S}_{i}^{2}=4\left[\begin{array}{cccc}
2 & 0 & 0 & 0\\
0 & 1 & 1 & 0\\
0 & 1 & 1 & 0\\
0 & 0 & 0 & 2
\end{array}\right],\label{eq:non-diag-S2}
\end{equation}
rewritten the Hamiltonian in terms of operators $S_{i}^{z}$ and $\boldsymbol{S}_{i}^{2}$
we have an Ising-Heisenberg chain model whose Hamiltonian is given
by $H=H_{0}+H'$, with $H_{0}=(2J^{\prime}+\Delta^{\prime})N$. Therefore,
the transformed Hamiltonian reduce to
\begin{alignat}{1}
H'= & -\sum_{i=1}^{N}\left\{ \tfrac{J^{\prime}}{4}(\boldsymbol{S}_{i}^{2}+\boldsymbol{S}_{i+1}^{2})+\tfrac{\Delta^{\prime}-J^{\prime}}{4}\left[\left(S_{i}^{z}\right)^{2}+\left(S_{i+1}^{z}\right)^{2}\right]\right.\nonumber \\
 & \left.+J(\boldsymbol{\sigma}_{c,i},\boldsymbol{\sigma}_{d,i})_{\Delta}+J_{0}\left(S_{i}^{z}\sigma_{c,i}^{z}+\sigma_{d,i}^{z}S_{i+1}^{z}\right)\right\} .\label{eq:Ham-eff}
\end{alignat}

We can observe from eq.\eqref{eq:non-diag-S2} the matrix has a $2\times2$
block matrix; hence, we can diagonalize this block matrix. It is interesting
to note that the matrix $S_{i}^{z}$ is still diagonal in the new
basis, due to corresponding $2\times2$ block matrix be null. Then,
this means we can simultaneously diagonalize both matrices $S_{i}^{z}$
and $\boldsymbol{S}_{i}^{2}$. Recall those $S_{i}^{z}$ and $\boldsymbol{S}_{i}^{2}$
are commutative operators. So, the diagonal matrices are expressed
by
\begin{equation}
\boldsymbol{S}_{i}^{2}=\left[\begin{array}{cccc}
8 & 0 & 0 & 0\\
0 & 8 & 0 & 0\\
0 & 0 & 0 & 0\\
0 & 0 & 0 & 8
\end{array}\right]\quad\text{and}\quad S_{i}^{z}=\left[\begin{array}{cccc}
2 & 0 & 0 & 0\\
0 & 0 & 0 & 0\\
0 & 0 & 0 & 0\\
0 & 0 & 0 & -2
\end{array}\right],
\end{equation}
and whose corresponding eigenvector states are expressed as
\begin{align}
|\tau_{+1}\rangle= & |\begin{smallmatrix}+\\
+
\end{smallmatrix}\rangle,\\
|\tau_{0}\rangle= & \frac{1}{\sqrt{2}}\left(|\begin{smallmatrix}+\\
-
\end{smallmatrix}\rangle+|\begin{smallmatrix}-\\
+
\end{smallmatrix}\rangle\right),\\
|s_{0}\rangle= & \frac{1}{\sqrt{2}}\left(|\begin{smallmatrix}+\\
-
\end{smallmatrix}\rangle-|\begin{smallmatrix}-\\
+
\end{smallmatrix}\rangle\right),\\
|\tau_{-1}\rangle= & |\begin{smallmatrix}-\\
-
\end{smallmatrix}\rangle.
\end{align}

The effective Ising \textquotedbl{}spin\textquotedbl{} in Hamiltonian
\eqref{eq:Ham-eff} can be understood as a composition of one triplet
state and one singlet state.

This transformation is possible because, the local gauge symmetry
is only satisfied by Hamiltonian \eqref{eq:Ham-orig} when the Ising
couplings $ac$ ($ad$) and $bc$($bd$) are identical.

\subsection{The zero temperature phase diagram}

In order to analyze the phase diagram of the orthogonal-dimer plaquette,
we assume the particular case $J^{\prime}=J$ and $\Delta^{\prime}=\Delta$,
following the parameter used in the literature\cite{koga,koga02,schu}. 

The dimer-plaquette Ising-Heisenberg chain described by the Hamiltonian
\eqref{eq:Ham-orig} exhibit five ground states. These states are
expressed as
\begin{align}
|FM\rangle= & \prod_{i=1}^{N}|\begin{smallmatrix}+\\
+
\end{smallmatrix}\rangle_{i}\otimes|++\rangle_{i},\label{FM}\\
|AFM\rangle= & \prod_{i=1}^{N}|\begin{smallmatrix}+\\
+
\end{smallmatrix}\rangle_{i}\otimes|--\rangle_{i}\label{AFM},
\end{align}

For the ferromagnetic (FM) phase and  antiferromagnetic (AFM) phase the corresponding eigenvalues are given by 
\begin{align*}
E_{FM}= & -2J^{\prime}-2\Delta^{\prime}-\Delta-4J_{0},\\
E_{AFM}= & -2J^{\prime}-2\Delta^{\prime}-\Delta+4J_{0}.
\end{align*}

The other phase are composed by dimers in triplet-triplet (TT) phase
and triplet-singlet (TS) phase, 
\begin{align}
|TT\rangle= & \prod_{i=1}^{N}|\tau_{0}\rangle_{i}\otimes\frac{1}{\sqrt{2}}\left(|+-\rangle_{i}+|-+\rangle_{i}\right),\label{eq:TT}\\
|TS\rangle= & \prod_{i=1}^{N}|\tau_{0}\rangle_{i}\otimes\frac{1}{\sqrt{2}}\left(|+-\rangle_{i}-|-+\rangle_{i}\right),\label{eq:TS}
\end{align}
where the corresponding eigenvalues become 
\begin{align*}
E_{TT}= & \Delta-2J-4J^{\prime},\\
E_{TS}= & \Delta+2{\it J}.
\end{align*}
It is worthy to mention that the state given by eqs.\eqref{eq:TT}
and \eqref{eq:TS} becomes frustrated state only when $J=0$ (Ising
limit).

Finally, the dimer-antiferromagnetic (DFA) state can be expressed
by
\begin{align}
|DFA\rangle= & \prod_{i=1}^{N/2}|\begin{smallmatrix}+\\
+
\end{smallmatrix}\rangle_{i}\otimes|\eta_{2,-2}\rangle_{i}\otimes|\begin{smallmatrix}-\\
-
\end{smallmatrix}\rangle_{i}\otimes|\eta_{2,-2}\rangle,
\end{align}
with
\begin{equation}
|\eta_{2,-2}\rangle=\frac{\left(|+-\rangle_{i}+\vartheta|-+\rangle_{i}\right)}{\sqrt{1+\vartheta^{2}}},
\end{equation}
and
\begin{equation}
\vartheta=\frac{\sqrt{4J_{0}^{2}+J^{2}}+2J_{0}}{J}.
\end{equation}

Hereafter the corresponding dimer-antiferromagnetic eigenvalue becomes
\begin{align*}
E_{DFA}= & -2J^{\prime}-2\Delta^{\prime}+\Delta-2\sqrt{4J_{0}^{2}+J^{2}}.
\end{align*}

\begin{figure}
\begin{centering}
\includegraphics[scale=0.22]{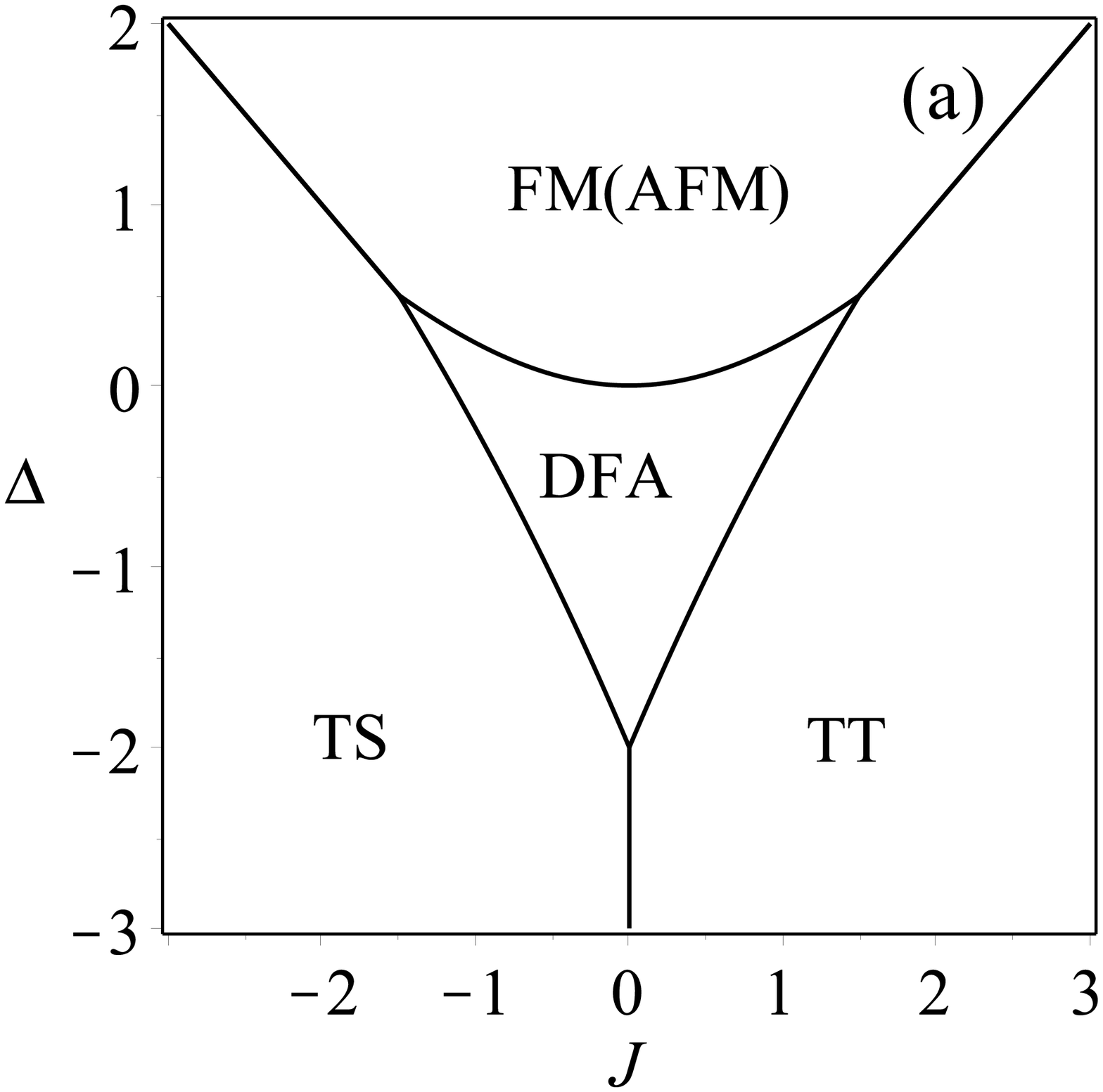}\includegraphics[scale=0.22]{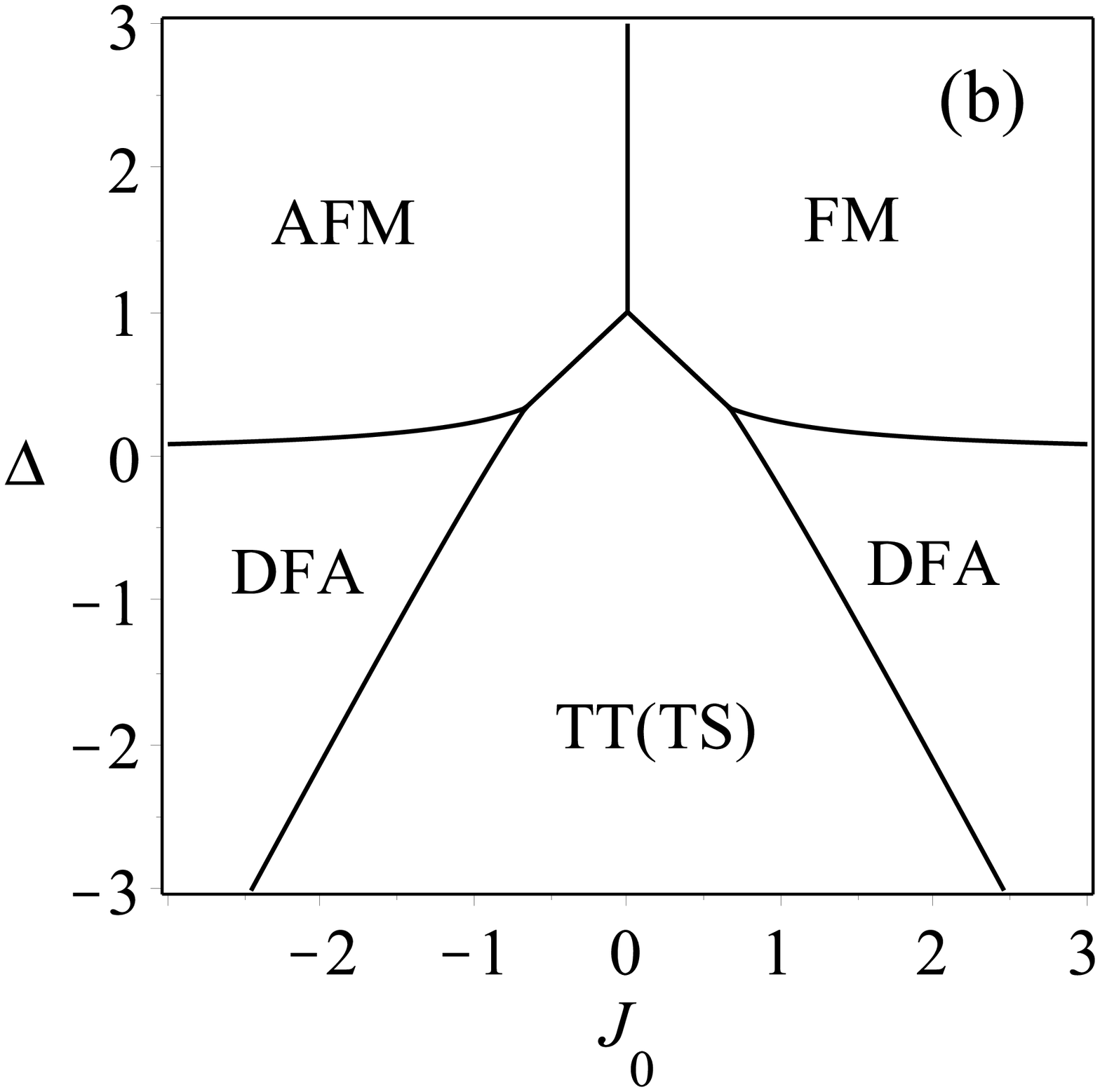}
\par\end{centering}

\caption{\label{fig:Phase-diagram} (a) Phase diagram at zero temperature as
a dependency of $J$ and $\Delta$ for a fixed value of $J_{0}=1$.
(b) Phase diagram at zero temperature as a dependency of $J_{0}$
and $\Delta$ for a fixed value of $J=1$.}
\end{figure}
In order to display the phase diagram at zero temperature, we plot
$J$ versus $\Delta$, assuming fixed value for $J_{0}=1(-1)$ FM(AFM)
respectively. As shown in figure \ref{fig:Phase-diagram}(a). For
$\Delta<-2$, there is a boundary between TS and TT state at $J=0$,
whereas for $-2\leqslant\Delta\leqslant0.5$ the state TS (TT) is
limited by DFA region whose boundary is given by $\Delta=2|J|-\sqrt{4+J^{2}}$,
when $J>0$ ($J<0$) respectively. While for $\Delta\geqslant0.5$
the state TS (TT) is limited by FM(AFM) region $\Delta=|J|-1$. Furthermore,
the boundary between DFA and FM(AFM) region is represented by the
curve $\Delta=\sqrt{4+J^{2}}-2$.

While in figure \ref{fig:Phase-diagram}(b), we observe the phase
diagram from another viewpoint, $\Delta$ against $J_{0}$, assuming
fixed $J=1$. In this picture the TS and TT phase is displayed in
same region, while AFM and FM states are illustrated for $J_{0}>0$
($J_{0}<0$) respectively. The boundary between FM(AFM) and TT(TS)
is given by $\Delta=1-|J_{0}|$  when $|J_{0}|<2/3$. Whereas the DFA
region appears when $|J_{0}|>2/3$, which is limited by TT(TS) region
and this boundary curve becomes $\Delta=2-\sqrt{4J_{0}^{2}+1}$, and
finally the boundary between FM and AFM regions is described by the curve
$\Delta=\sqrt{4J_{0}^{2}+1}-2|J_{0}|$.

\section{The thermodynamics of the model}

The method to be used will be the decoration transformation proposed
in early 50 decade by Syozi \cite{Syozi} and Fisher \cite{Fisher}.
Afterward this approach was the subject of study in reference \cite{phys-A-09},
for the case of multi-spins. Similar generalization was developed
by Stre\v{c}ka \cite{strecka pla} for the hybrid system (e.g. Ising-Heisenberg).
Another interesting variant of this approach also was discussed previously in reference \cite{JPA-11},
where  a direct transformation was proposed instead of several step
by step transformation. Consequently, the decoration
transformation approach is widely used to solve spins models, besides,
the decoration transformation approach can also be applied to electron
coupling system such has been applied for the case of spinless fermion
on diamond structure\cite{spinless}, and as well as for Hubbard model
in the quasi-atomic limit\cite{hubbard}.

In order to study the thermodynamics of the dimer-plaquette Ising-Heisenberg
chain, we will use the decoration transformation proposed in reference
\cite{JPA-11} together with the usual transfer matrix technique \cite{baxter-book}.
So, let us start considering the partition function as follow
\begin{equation}
\mathcal{Z}_{N}=\mathrm{e}^{-\beta H_{0}}\mathrm{tr}\left(\prod_{i=1}^{N}\mathrm{e}^{-\beta H'}\right),
\end{equation}
where $\beta=1/kT$, with $k$ being the Boltzmann constant and $T$
is the absolute temperature, and assuming $H'$ is given by eq.\eqref{eq:Ham-eff}.

The effective model can be solved by the usual transfer matrix approach\cite{baxter-book}.
Inasmuch as the transfer matrix of the Hamiltonian \eqref{eq:Ham-eff}
is reduced to
\begin{equation}
\boldsymbol{T}=\left[\begin{array}{cccc}
w_{1,1} & w_{1,0}{x^{\prime}}^{2} & w_{1,0} & w_{1,-1}\\
w_{1,0}{x^{\prime}}^{2} & w_{0,0}{x^{\prime}}^{4} & w_{0,0}{x^{\prime}}^{2} & w_{1,0}{x^{\prime}}^{2}\\
w_{1,0} & w_{0,0}{x^{\prime}}^{2} & w_{0,0} & w_{1,0}\\
w_{1,-1} & w_{1,0}{x^{\prime}}^{2} & w_{1,0} & w_{1,1}
\end{array}\right],
\end{equation}
with
\begin{alignat}{1}
w_{1,1}= & {x^{\prime}}^{2}{z^{\prime}}^{2}\left[z\left(y^{4}+y^{-4}\right)+z^{-1}(x^{2}+x^{-2})\right],\\
w_{1,0}= & {x^{\prime}}{z^{\prime}}\left[z\left(y^{2}+y^{-2}\right)+z^{-1}(x_{1}^{2}+x_{1}^{-2})\right],\\
w_{1,-1}= & {x^{\prime}}^{2}{z^{\prime}}^{2}\left[2z+z^{-1}\left(x_{2}^{2}+x_{2}^{-2}\right)\right],\\
w_{0,0}= & 2z+z^{-1}\left(x^{2}+x^{-2}\right),
\end{alignat}
where we have introduced the following notations $x=\mathrm{e}^{\beta J}$
, $z=\mathrm{e}^{\beta\Delta}$, ${x^{\prime}}=\mathrm{e}^{\beta J^{\prime}}$
, ${z^{\prime}}=\mathrm{e}^{\beta\Delta^{\prime}}$ and $y=\mathrm{e}^{\beta J_{0}}$.
Furthermore, we define also the following exponentials $x_{1}=\mathrm{e}^{\beta\sqrt{J^{2}+J_{0}^{2}}}$
and $x_{2}=\mathrm{e}^{\beta\sqrt{J^{2}+4J_{0}^{2}}}$.

In order to diagonalize the transfer matrix $\boldsymbol{T}$, we
perform the determinant of the transfer matrix, which is expressed
as follow
\begin{equation}
\lambda(\lambda^{2}-a_{1}\lambda+a_{0})(\lambda-w_{1,1}+w_{1,-1})=0,\label{eq:cubic-eq}
\end{equation}
here, the coefficients of the quadratic term eq.\eqref{eq:cubic-eq}
is given by
\begin{align}
a_{1}= & w_{1,1}+w_{0,0}(1+{x^{\prime}}^{4})+w_{1,-1},\label{eq:a1}\\
a_{0}= & \left(1+{x^{\prime}}^{4}\right)\left[w_{0,0}\left(w_{1,1}+w_{1,-1}\right)-2w_{1,0}^{2}\right],\nonumber 
\end{align}

Therefore, the eigenvalues can be expressed as 
\begin{align}
\lambda_{0}= & \tfrac{1}{2}\left(a_{1}+\sqrt{a_{1}^{2}-4a_{0}}\right),\\
\lambda_{1}= & \tfrac{1}{2}\left(a_{1}-\sqrt{a_{1}^{2}-4a_{0}}\right),\\
\lambda_{2}= & {x^{\prime}}^{2}{z^{\prime}}^{2}\left[z\left(y^{2}-y^{-2}\right)^{2}+\frac{\left(x^{2}+x^{-2}-x_{2}^{2}-x_{2}^{-2}\right)}{z}\right],\\
\lambda_{3}= & 0.
\end{align}

In the thermodynamic limit, the free energy per unit cell is given
only by the largest eigenvalue of transfer matrix, and for our case,
we can easily verify the first eigenvalues $\lambda_{0}$ become always
the largest one. Thus, the free energy can be expressed by the relation
below
\begin{equation}
f=2J^{\prime}+\Delta^{\prime}-\frac{1}{\beta}\ln\left(\lambda_{0}\right).\label{eq:free-e}
\end{equation}

The first term of the free energy is a trivial constant energy, obtained
during the local gauge transformation, which is irrelevant for thermodynamis
quantities. Note that the free energy is valid for arbitrary value
of $J^{\prime}$ and $\Delta^{\prime}$, although here we consider
only a particular case $J^{\prime}=J$ and $\Delta^{\prime}=\Delta$,
following the parameter used in the literature\cite{koga,koga02,schu}.
Therefore, any additional properties can be obtained straightforwardly
from eq.\eqref{eq:free-e}.

\subsection{Entropy and specific heat}

\begin{figure}
\centering{}\includegraphics[scale=0.3]{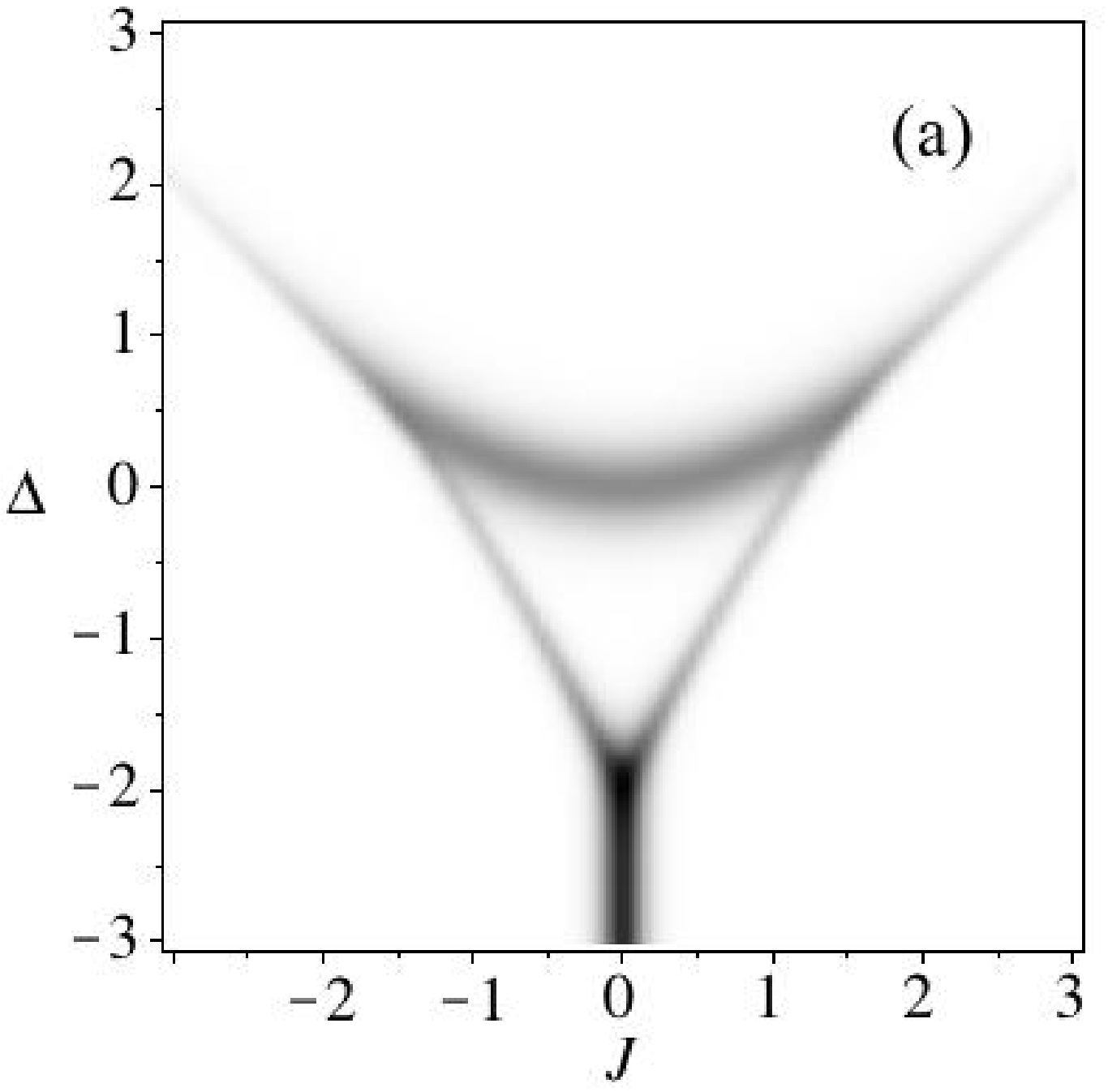}\includegraphics[scale=0.3]{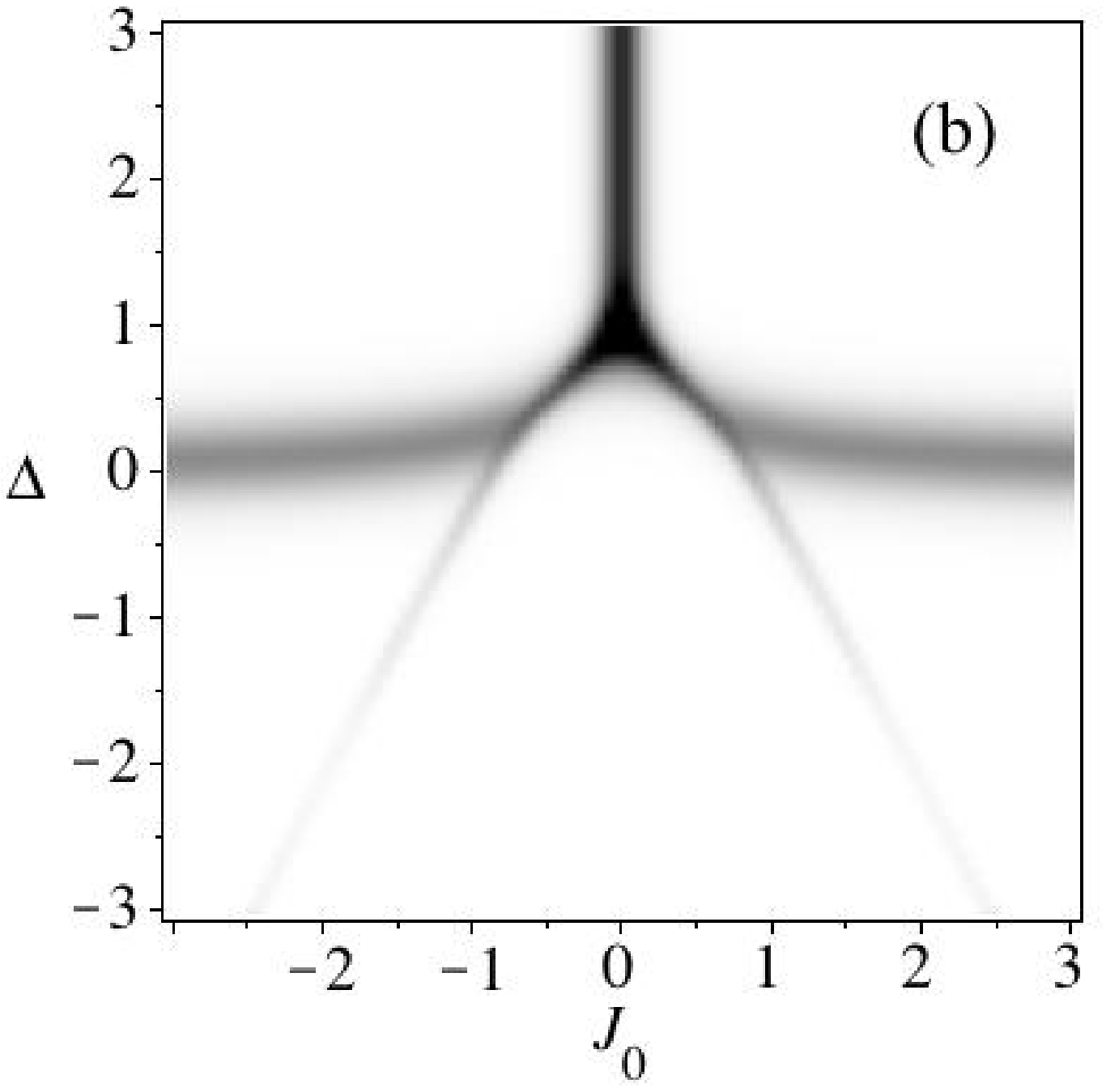}\caption{\label{fig:entropy}Density plot of entropy, darkest region corresponds
to higher entropy. (a) Entropy as a dependency of $J$ and $\Delta$
for a fixed value of $J_{0}=1$. (b) Entropy as a dependency of $J_{0}$
and $\Delta$ for a fixed value of $J=1$.}
\end{figure}

In order to accomplish with our discussion concerning to the thermodynamics
properties. Let us illustrate the entropy ($\mathcal{S}=-\partial f/\partial T$)
in fig. \ref{fig:entropy}(a) for the low temperature limit as a function
of $\Delta$ and $J$, whereas, in fig.\ref{fig:entropy}(b) the entropy
is illustrated as a function of $\Delta$ and $J_{0}$. Dark region
corresponds to higher entropy, while the white region corresponds
to zero entropy. However, as soon as the temperature decreases the
entropy leads to zero, unless, for $J=0$ and $\Delta\leqslant-2$
(pure Ising chain) the dark line will remain when temperature decreases,
leading to a residual entropy $\mathcal{S}_{0}=2\ln(2)$, which illustrates
the geometric frustration region of orthogonal plaquette Ising chain.

\begin{figure}
\centering{}\includegraphics[scale=0.4]{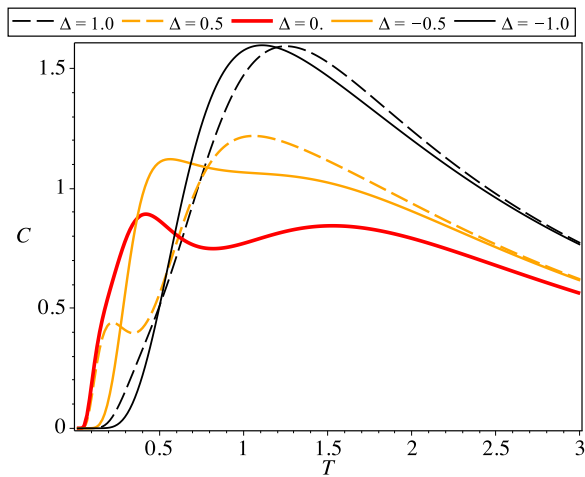}\caption{\label{fig:spcfic-ht}(Color online) Specific heat as a temperature
dependency, for several values of anisotropic parameter assuming fixed values $J=1.0$ and $J_0=01.0$.}
\end{figure}

Finally, we discuss also another interesting thermodynamic quantity
called the specific heat ($C=-T\partial^{2}f/\partial T^{2}$). In
figure \ref{fig:spcfic-ht}, we plot the specific heat as a function
of the temperature $T$ for a fixed coupling parameter $J_{0}=1.0$
and a fixed value of $J=1.0$. By a solid line is represented the
specific heat curve when $\Delta=0$, $0.5$ and $1.0$, while, for
dashed line we illustrate the specific heat for $\Delta=-0.5$, and
$-1.0$. For anisotropic parameter $\Delta=\pm0.5$ and $0$, the
plot illustrates a small anomalous peak due to the zero temperature
phase transition influence, although for $\Delta=\pm1.0$ this anomalous
peak disappear, because there is no zero temperature phase transition in the neighborhood.

\subsection{Pair correlation function}

The nearest site correlation function between dimers can be obtained
using a derivative of the free energy given by eq.\eqref{eq:free-e}.
Then, in order to perform the derivative of free energy, we need to
assume the parameter $J^{\prime}$ and $J$ as independent parameter,
similarly $\Delta^{\prime}$ and $\Delta$ also are considered as
independent parameters. Despite for physical quantities, the term
$2J^{\prime}+\Delta^{\prime}$ in the free energy be irrelevant, this
term cannot be neglected when we calculate the correlation function
as expressed below:
\begin{alignat}{1}
\langle\sigma_{a}^{x}\sigma_{b}^{x}\rangle= & -\frac{1}{2}\frac{\partial f}{\partial J^{\prime}}=-1+\frac{x^{\prime}}{2\lambda_{0}}\frac{\partial\lambda_{0}}{\partial x^{\prime}},\label{sabxx}\\
\langle\sigma_{a}^{z}\sigma_{b}^{z}\rangle= & -\frac{\partial f}{\partial\Delta^{\prime}}=-1+\frac{z^{\prime}}{\lambda_{0}}\frac{\partial\lambda_{0}}{\partial z^{\prime}},\\
\langle\sigma_{c}^{x}\sigma_{d}^{x}\rangle= & -\frac{1}{2}\frac{\partial f}{\partial J}=\frac{x}{2\lambda_{0}}\frac{\partial\lambda_{0}}{\partial x},\\
\langle\sigma_{c}^{z}\sigma_{d}^{z}\rangle= & -\frac{\partial f}{\partial\Delta}=\frac{z}{\lambda_{0}}\frac{\partial\lambda_{0}}{\partial z}.\label{scdzz}
\end{alignat}

Here, we present the relation of the correlation function in terms
of the largest eigenvalue of the transfer matrix $\lambda_{0}$.

\begin{figure}
\centering{}\includegraphics[scale=0.25]{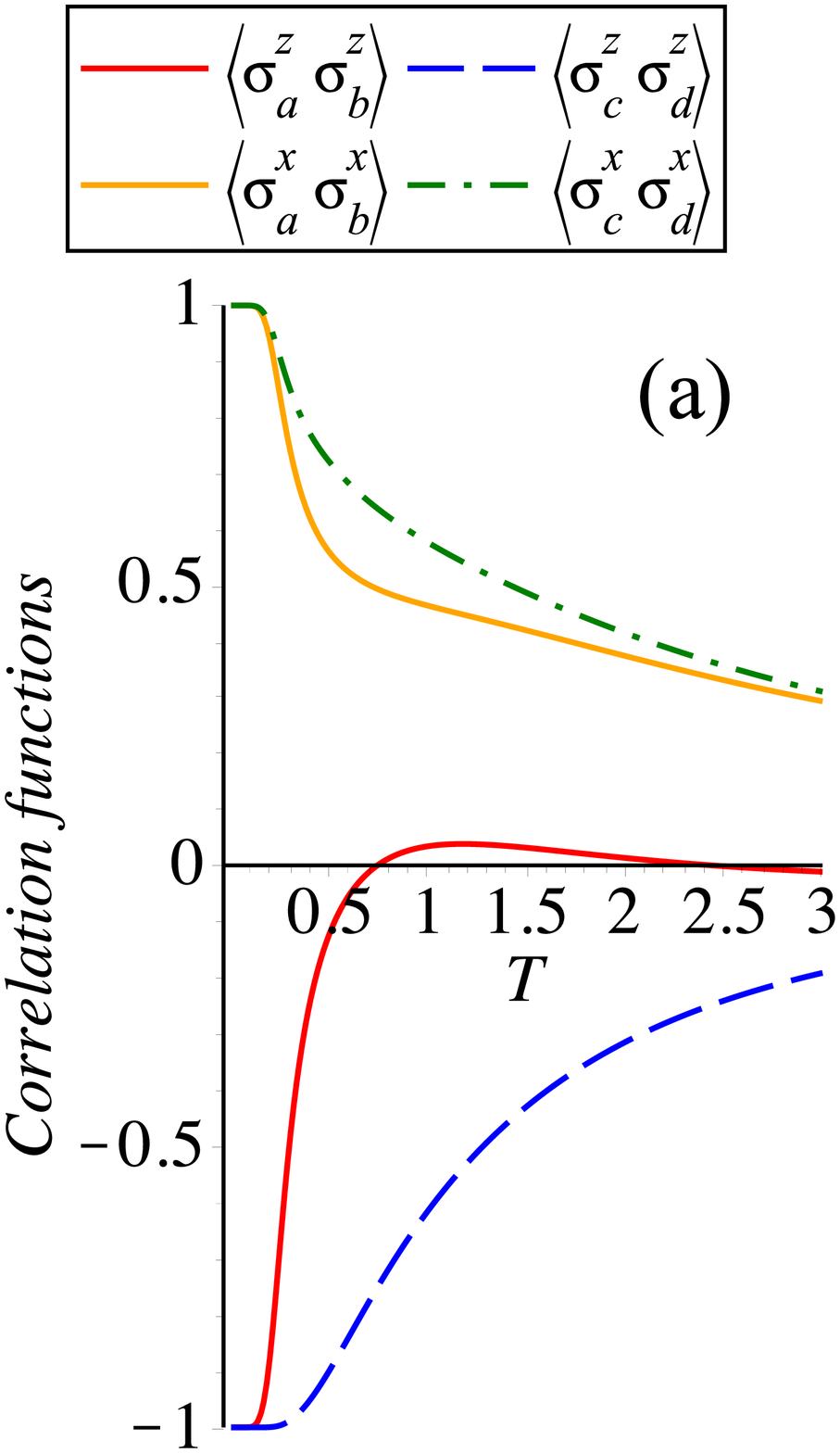}\includegraphics[scale=0.25]{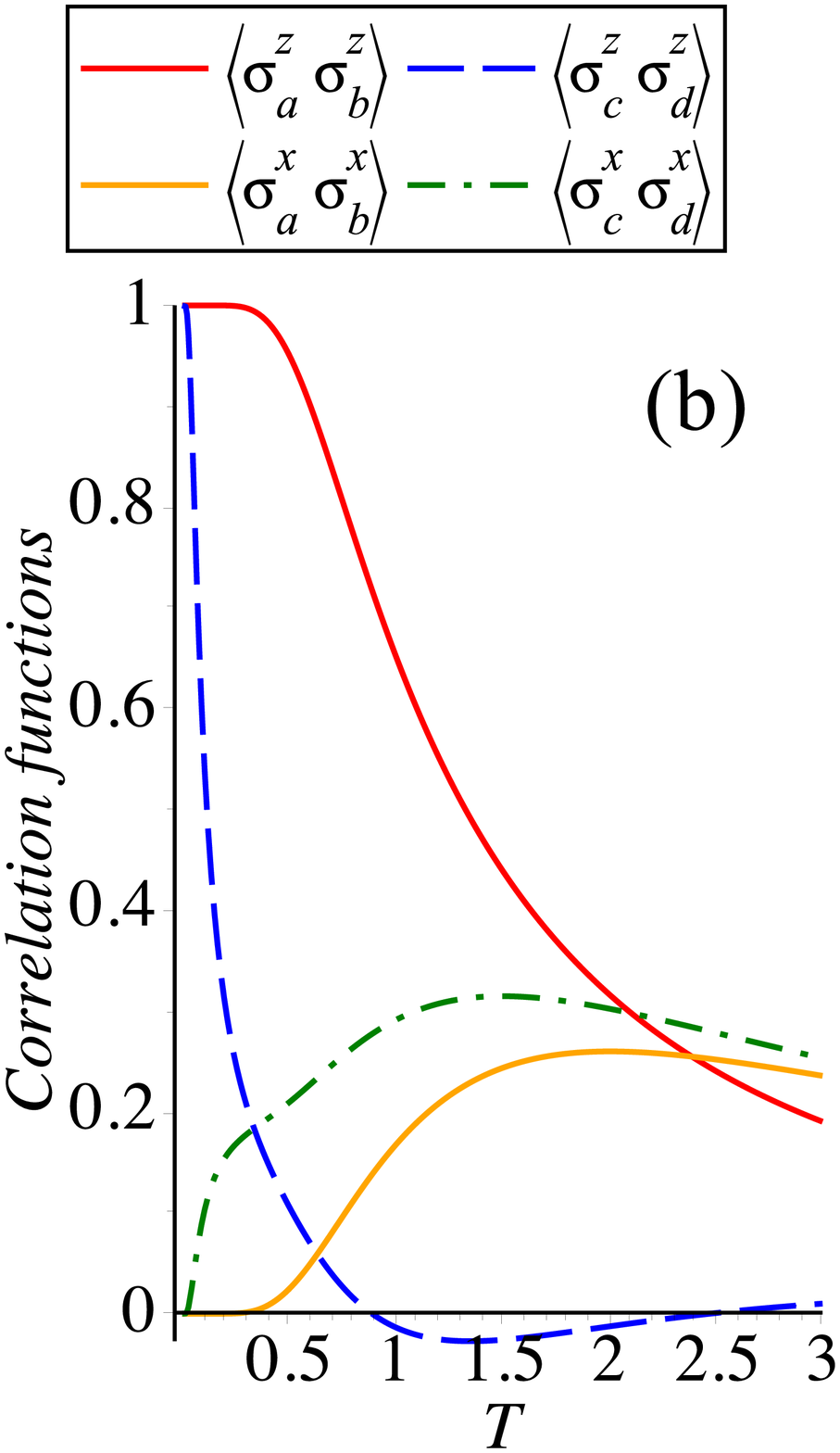}\caption{\label{fig:correlation}Correlation functions for $ab$-dimer and $cd$-dimer defined by eqs. (\ref{sabxx}-\ref{scdzz}), assuming fixed parameter $J=1.0$ and $J_0=1.0$. (a) For the case of anisotropic parameter $\Delta=-0.3$. (b) And for the case of anisotropic parameter $\Delta=0.3$. }
\end{figure}

 In order to accomplish our analysis concerning to the correlation function we display in fig.\ref{fig:correlation} the correlation function given by eqs. (\ref{sabxx}-\ref{scdzz}), as a function of temperature, assuming convenientely fixed parameters $J_0=1.0$ and $J=1.0$  close to the phase transition illustrated in fig.\ref{fig:Phase-diagram}. Thus, in fig.\ref{fig:correlation}(a) we display for the case of anisotropic parameter $\Delta=-0.3$, and we observe how the correlation function at low temperature behaves quite similarly for both the $ab$-dimer and the $cd$-dimer; for example the $z$-component of pair correlation are negative because the ground state energy is TT(TS), while the $xy$-component are positive.  
 However as soon the temperature increases the  difference becomes relevant. It is worth to emphasize that the correlation function   $\langle\sigma_{a}^{z}\sigma_{b}^{z}\rangle$ turn positive at finite temperature $T\approx0.7$ and for higher temperature once again the correlation function  is negative at a temperature of around $T\approx2.5$ becoming a small negative amount. Similarly we also plot  in fig.\ref{fig:correlation}(b) for anisotropic parameter $\Delta=0.3$, the $ab$-dimer and $cd$-dimer behaves quite similarly in the low temperature limit, the $z$-componet of pair correlation are positive because it corresponds to FM(AFM) region in agreement with eq.\eqref{FM} and \eqref{AFM}, while  the $xy$-componet   correlation function is zero. Nevertheless, for higher temperature this difference is significant, the $\langle\sigma_{a}^{x}\sigma_{b}^{x}\rangle$ and $\langle\sigma_{c}^{x}\sigma_{d}^{x}\rangle$ have a maximum at finite temperature,  while the  $\langle\sigma_{a}^{z}\sigma_{b}^{z}\rangle$ is a monotonic decreasing function and $\langle\sigma_{c}^{z}\sigma_{d}^{z}\rangle$ decreases faster  becoming negative at $T\approx 0.9$ and for higher temperature at around $T\approx2.4$ the $\langle\sigma_{c}^{z}\sigma_{d}^{z}\rangle$ correlation function becomes a positive quantity.

\section{Thermal entanglement}

Another interesting property we consider in this work, will be the
quantum entanglement of the Ising-Heisenberg orthogonal dimer plaquette
model. As a measure of entanglement for two arbitrary mixed states
of dimers, we use the quantity called concurrence\cite{wooters},
which is defined in terms of reduced density matrix $\rho$ of two
mixed states
\begin{equation}
\mathcal{C}(\rho)=\max\{{0,2\Lambda_{max}-\text{{tr}}\sqrt{R}}\},
\end{equation}
assuming
\begin{equation}
R=\rho\sigma^{y}\otimes\sigma^{y}\rho^{*}\sigma^{y}\otimes\sigma^{y},
\end{equation}
where $\Lambda_{max}$ is the largest eigenvalue of the matrix $\sqrt{R}$
and $\rho^{*}$ represent the complex conjugate of matrix $\rho$,
with $\sigma^{y}$ being the Pauli matrix.

For the case of infinite chain, the reduced density operator elements\cite{bukman}
could be expressed in terms of the correlation function between two
entangled particles\cite{amico}, consequently the concurrence between
$ab$-dimer becomes
\begin{equation}
\mathcal{C}_{ab}=\max\{0,|\langle\sigma_{a}^{x}\sigma_{b}^{x}\rangle|-\frac{1}{2}|1+\langle\sigma_{a}^{z}\sigma_{b}^{z}\rangle|\},\label{eq:Cnr-ab}
\end{equation}
similarly the concurrence between $cd$-dimer reads
\begin{equation}
\mathcal{C}_{cd}=\max\{0,|\langle\sigma_{c}^{x}\sigma_{d}^{x}\rangle|-\frac{1}{2}|1+\langle\sigma_{c}^{z}\sigma_{d}^{z}\rangle|\}.\label{eq:Cnr-cd}
\end{equation}

Surely, we can obtain also equivalent result using the approach described
in reference \cite{spra}.

It is worthy to mention that the zero temperature entanglement for
$ab$-dimer is maximally entangled in TS and TT region $\mathcal{C}_{ab}=1$,
while, in all other region the $ab$-dimer becomes untangled ($\mathcal{C}_{ab}=0$).
Whereas the $cd$-dimer also is maximally entangled in same (TS and
TT) region $\mathcal{C}_{cd}=1$; additionally the DFA region now also
becomes an entangled region whose concurrence is given by $\mathcal{C}_{cd}=\frac{|J|}{\sqrt{4J_{0}^{2}+J^{2}}}$
and it depends of $J$ and $J_{0}$. It is worth to notice, when $J=0$
this region becomes untangled, which is perfectly coherent since the
model reduces to orthogonal plaquette Ising chain. Another special
possibility is when $J_{0}=0$, the orthogonal dimer plaquette reduces
simply to independent entangled dimers. While the FM and AFM region
are untangled $\mathcal{C}_{cd}=0$.

\begin{figure}
\includegraphics[scale=0.4]{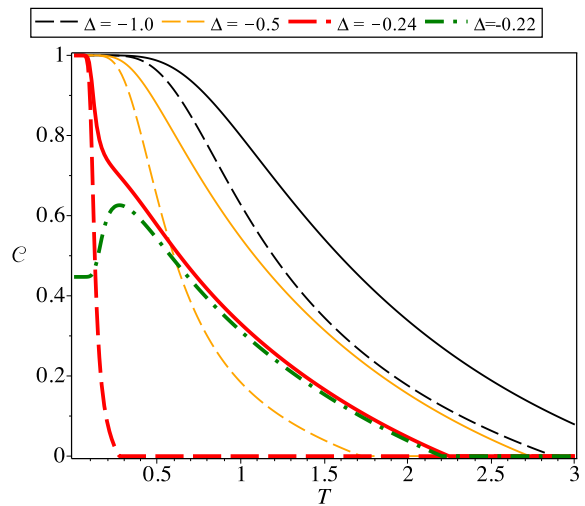}\caption{\label{fig:Concurr-T}(Color online) Concurrence for several values
of anisotropic parameter as a function of $T$, assuming fixed $J_{0}=1$ and $J=1$. Dashed line corresponds
for $ab$-dimer while the solid line represents $cd$-dimer.}
\end{figure}

In figure \ref{fig:Concurr-T} is plotted the concurrence $\mathcal{C}$
as a function of the temperature: for $\Delta=-1$ the dashed and
solid line corresponds to $ab$-dimer and $cd$-dimer respectively.
However, the $ab$-dimer entanglement vanishes for $T\approx2.9$
while, for $cd$-dimer the entanglement vanishes at $T\approx4$.0.
For weaker anisotropic coupling ($\Delta$) this difference becomes
more significantly, say i.e. for $\Delta=-0.24$, the $cd$-dimer
(solid line) entanglement vanishes for $T\approx2.25$, whilst for
$ab$-dimer the entanglement vanishes at $T\approx0.28$ (dashed line).
For a bit weaker anisotropic parameter $\Delta=-0.22$, the $ab$-dimer
becomes untangled region for any temperature, while $cd$-dimer entanglement
vanishes at $T\approx2.24$.

\subsection{Threshold temperature}

The threshold temperature can be obtain when $ab$-dimer concurrence
becomes null ($|\langle\sigma_{a}^{x}\sigma_{b}^{x}\rangle|=\frac{1}{2}|1+\langle\sigma_{a}^{z}\sigma_{b}^{z}\rangle|$),
and similarly for $cd$-dimer, the concurrence will become null when
($|\langle\sigma_{c}^{x}\sigma_{d}^{x}\rangle|=\frac{1}{2}|1+\langle\sigma_{c}^{z}\sigma_{d}^{z}\rangle|$).

\begin{figure}
\centering{}\includegraphics[scale=0.4]{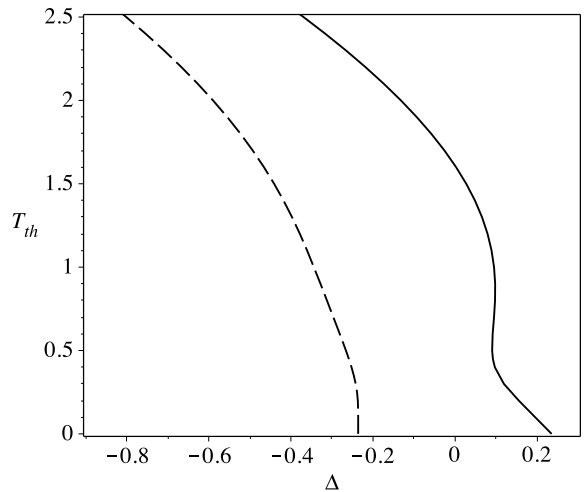}\caption{\label{fig:th-tem}Threshold temperature $T_{th}$ as a function of
$\Delta$, for fixed $J_{0}=1$ and $J=1$. Dashed line corresponds
for $ab$-dimer, while the solid line represents $cd$-dimer.}
\end{figure}

The threshold temperature is illustrated in figure \ref{fig:th-tem}
as a function of anisotropic parameter for fixed values of $J_{0}=1$
and $J=1$. The dashed line correspond for the $ab$-dimer threshold
temperature, and in the low temperature limit, the threshold temperature
leads to $T_{th}=2-\sqrt{5}$. Whereas the solid line represents the
$cd$-dimer threshold temperature; in the low temperature limit it
leads to $T_{th}=\sqrt{5}-2$. It is worthy to highlight that the
threshold temperature for $cd$-dimer exhibits a thin reentrance
around to $\Delta=0.1$.

\begin{figure}
\centering{}\includegraphics[scale=0.3]{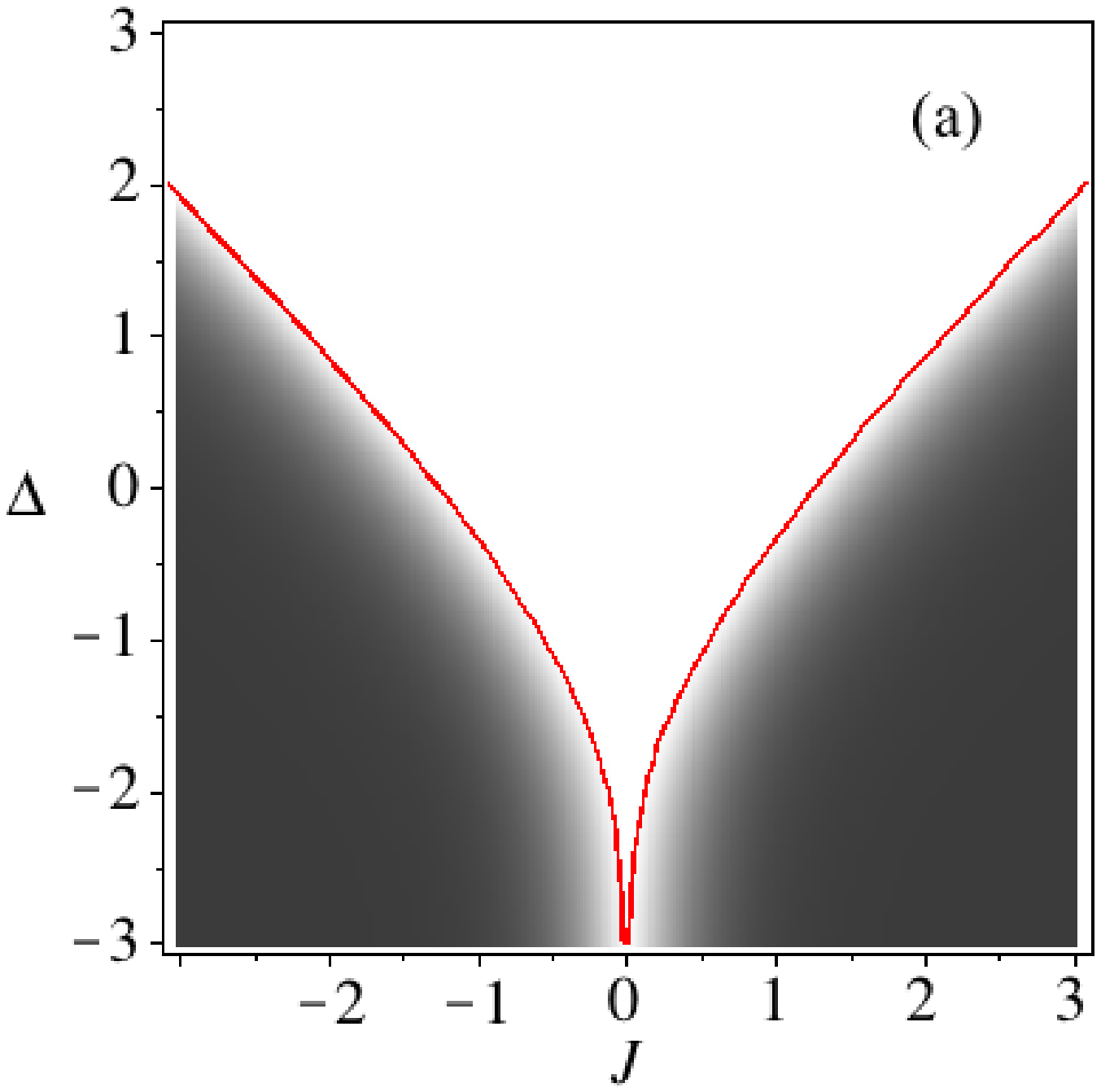}\includegraphics[scale=0.3]{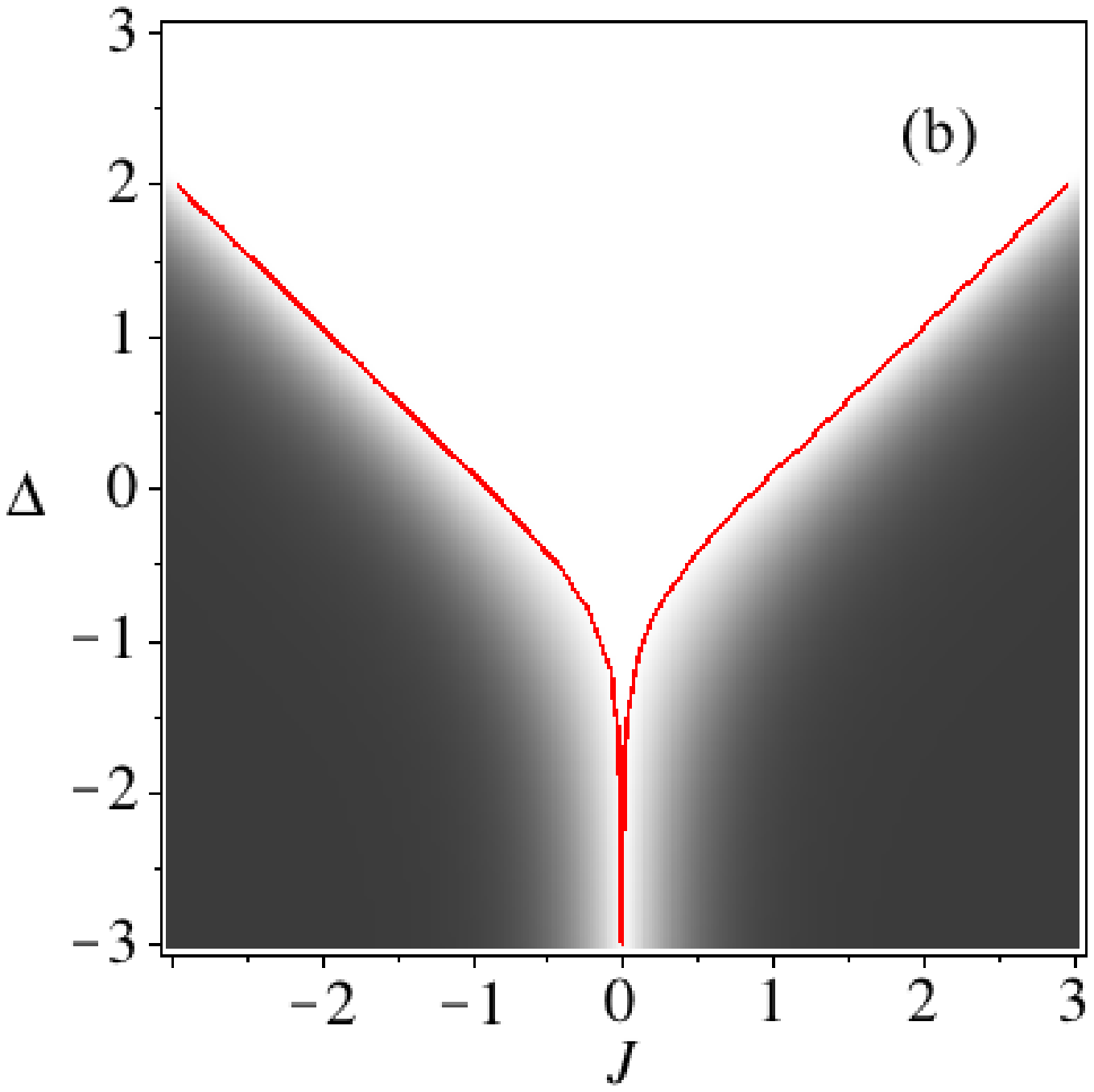}\caption{\label{fig:entag-JD}(Color online) Density plot of concurrence in units of $T_{th}$, darkest
region corresponds to higher entanglement. (a) Concurrence for $ab$-dimer
as a dependency of $J$ and $\Delta$ for a fixed value of $J_{0}=1$.
(b) Concurrence for $cd$-dimer as a dependency of $J$ and $\Delta$
for a fixed value of $J_{0}=1$.}
\end{figure}

Another way to display the entanglement phase diagram is in units
of threshold temperature as displayed in figure \ref{fig:entag-JD},
the boundary between entangled region and untangled region is given
by a solid (red) line. The black region correspond to maximally entangled
region, and white region corresponds to untangled region, while light
gray region corresponds to weak concurrence, similarly dark gray corresponds
to strong concurrence. The entangled phase diagram $J$ against $\Delta$
in units of $T_{th}$, behaves quite similar for both $ab$-dimer
and $cd$-dimer, although, for $ab$-dimer the untangled region appears
already significantly for $\Delta\geqslant-3$ and small values of
$J$, while for $cd$-dimer the untangled region becomes relevant
only for $\Delta\geqslant-1$ and small values of $J$. Certainly
this is in agreement with the zero temperature phase diagram displayed
in figure \ref{fig:Phase-diagram}(a).

\begin{figure}
\centering{}\includegraphics[scale=0.3]{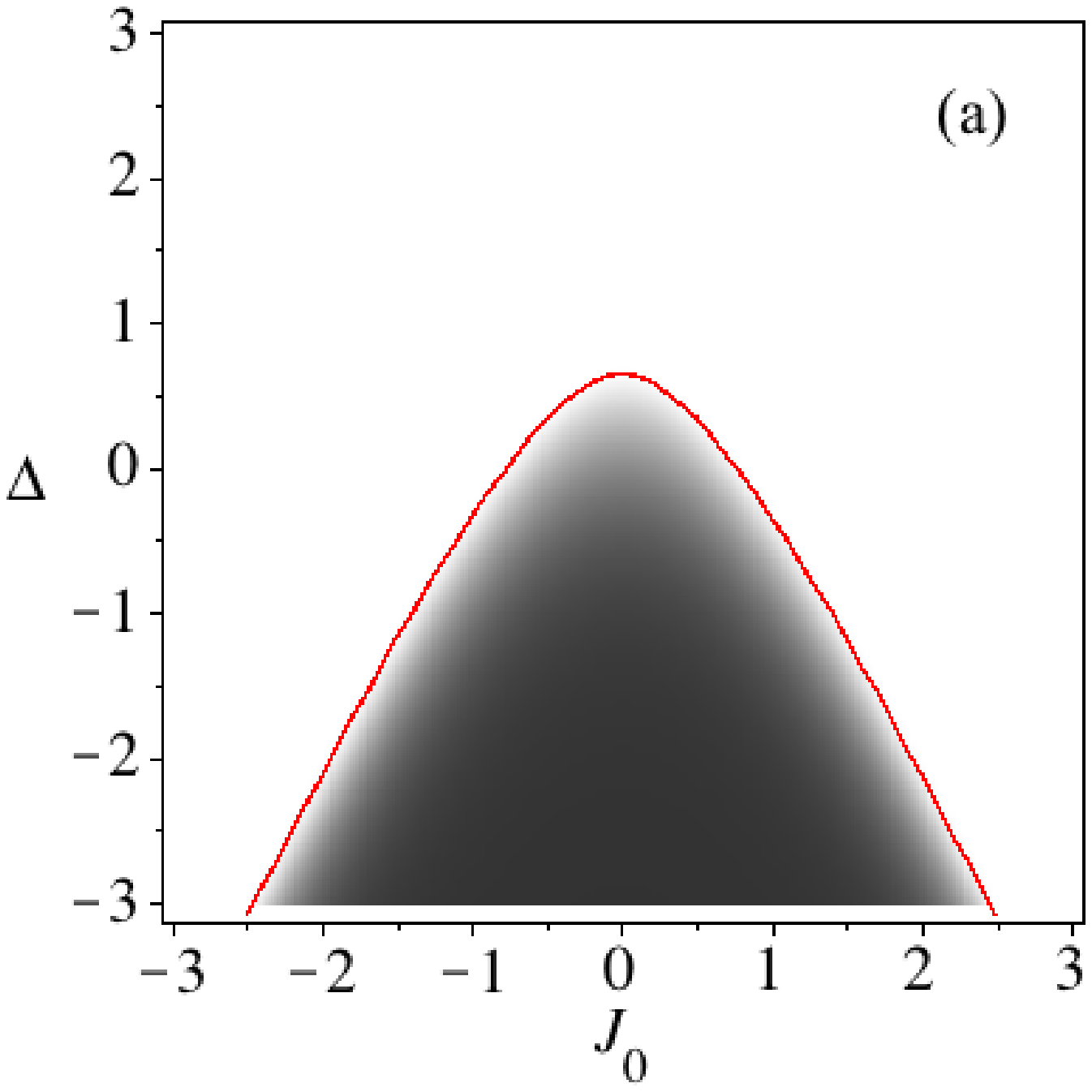}\includegraphics[scale=0.3]{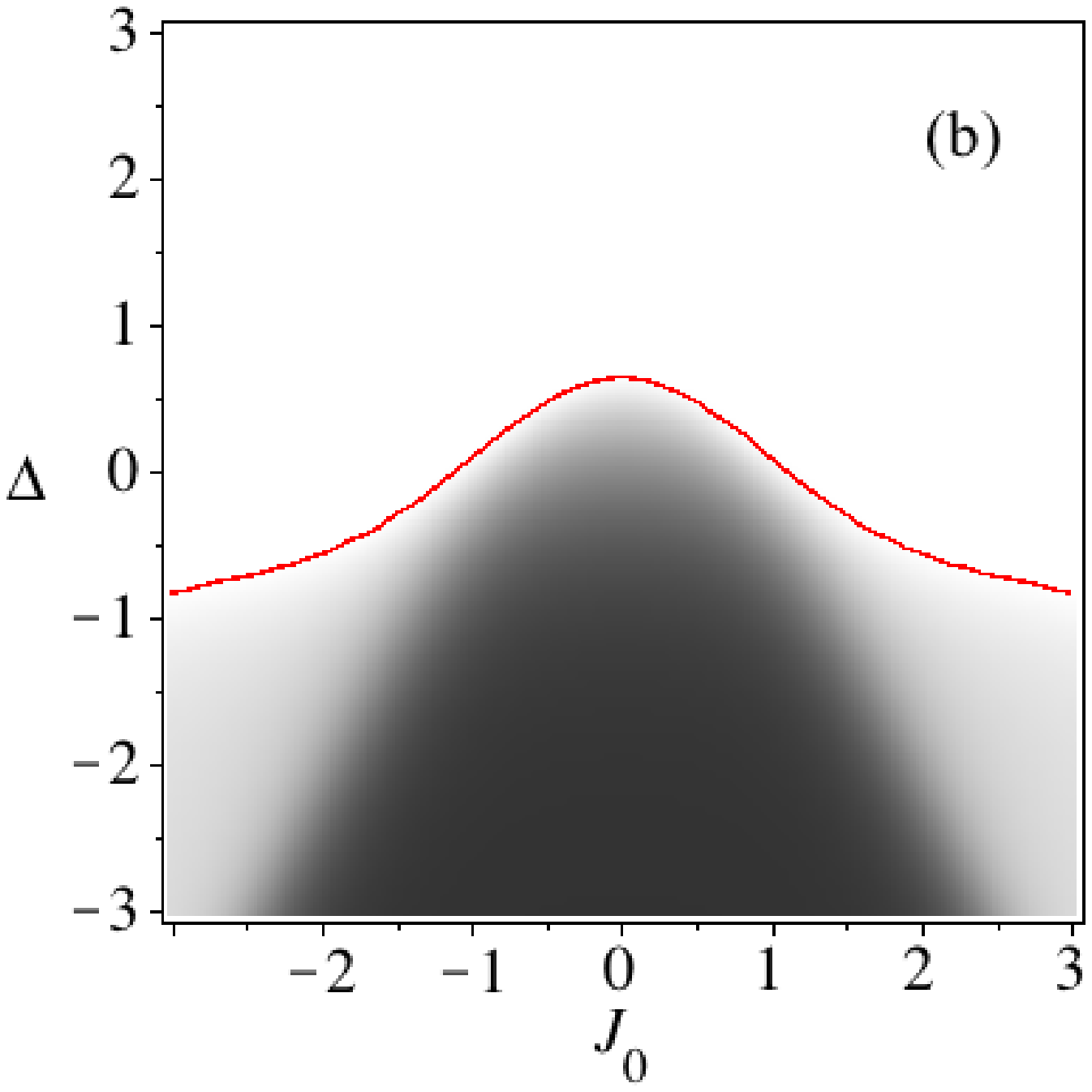}\caption{\label{fig:entag-J0D}Density plot of Concurrence in units of $T_{th}$, darkest re\cite{baxter-book}region
corresponds to higher entanglement. (a) Concurrence for $ab$-dimer
as a dependency of $J_{0}$ and $\Delta$ for a fixed value of $J=1$.
(b) Concurrence for $cd$-dimer as a dependency of $J_{0}$ and $\Delta$
for a fixed value of $J=1$.}
\end{figure}

Additional density plot of entangled phase diagram is displayed in
figure \ref{fig:entag-J0D} in terms of $\Delta$ and $J_{0}$, once
again in units of $T_{th}$, the phase diagrams for $ab$-dimer is
given in fig.\ref{fig:Phase-diagram}(a), and the phase diagram for
$cd$-dimer is illustrated in fig.\ref{fig:Phase-diagram}(b). Here,
we observe the phase diagram between $ab$-dimer and $cd$-dimer are
quite different, but still are closely related to the zero temperature
phase diagram illustrated in figure \ref{fig:Phase-diagram}(b). Although
the darkest (strongly entangled) region are very similar for both
dimers. 

\section{Conclusion}

In this work, we studied the dimer-plaquette Ising-Heisenberg chain,
assembled between plaquette edge also known as orthogonal dimer plaquette\cite{ivanov,richter,schu,koga,koga02,vadim12}.
Using the local gauge symmetry of this model, we are able to map onto
a simple spin-1 like Ising and spin-1/2 Heisenberg dimer model with
single effective ion anisotropy. Thereafter, this model can be solved
using the decoration transformation\cite{Syozi,Fisher,phys-A-09,JPA-11}
and transfer matrix approach. First we discuss the phase diagram at
zero temperature of this model, where we found five ground states
as illustrated in figure \ref{fig:Phase-diagram}, one ferromagnetic,
one antiferromagnetic, one triplet-triplet disordered and triplet-singlet
disordered phase, beside a dimer ferromagnetic-antiferromagnetic phase.
It is interesting to remark that, in the limit of pure Ising model
this exhibits a frustrated region. Furthermore,  the
thermodynamic properties are discussed such as entropy, specific heat as well as
the correlation function.  Additionally, using the nearest site
correlation function it is possible to analyze the pairwise thermal
entanglement for both $ab$-dimer and $cd$-dimer, as well as the
threshold temperature of the entangled region are discussed as a function
of the Hamiltonian parameters. There is some significant difference
between both dimers ($ab$-dimers and $cd$-dimers) which is in agreement
in the low temperature limit. However, for strong entanglement, both
dimers are quite similar. As a consequence of this difference, we
can illustrate one interesting result, regarding to the reentrance
type of threshold temperature for $cd$-dimer, despite for $ab$-dimer
there is no reentrance temperature.
\begin{acknowledgments}
H. G. P. thanks CAPES for fully financial support, while S. M. S.
and O. R. thank FAPEMIG and CNPq for partial financial support. O.
R. also thanks ICTP for partial financial support.\end{acknowledgments}

\end{document}